# Exploring domain adaptation for deep neural network trained QSM

JUAN LIU[1] and Kevin Koch[2]

[1]Yale University, New Haven, CT, United States, [2]Medical College of Wisconsin, Milwaukee, WI, United States

## Synopsis

**We aim to address the domain adaptation problem of neural networks for QSM reconstruction which are learned from synthetic data while applied on real data. To address the unsupervised domain adaption with unknown QSM label of real data, we apply domain-specific batch normalization layers in the convolutional neural networks while allowing them to share all other model parameters. The proposed method is evaluated on multiple orientation datasets and single-orientation QSM datasets. Compared with TKD, MEDI, and DL-based method first training on synthetic datasets then model-based fine-tuning on real datasets, the proposed method achieved the best performance.**

## Introduction

Quantitative susceptibility mapping (QSM) is an MRI technique that estimates tissue magnetic susceptibility. The generation of QSM requires solving a challenging ill-posed field-to-source inversion problem. Recently, several deep learning (DL) QSM techniques [1,2,3,4] have been proposed and demonstrated impressive performance. Due to the inherent non-existent ground-truth QSM references, these techniques used either COSMOS [5] maps or synthetic data for network training. Synthetic data is easy to generate for network training. However, the trained model from synthetic data often adapts poorly to in-vivo dataset due to domain shift. Here, we introduce an easy domain adaption technique using domain-specific batch normalization [6] to address this problem.

## Method

[Neural Network]
A modified U-net for 3D data processing was utilized for the dipole inversion, with inputs of local field and brain mask and output of QSM. For the training of the network, whole simulated QSM maps were generated using single brain COSMOS map from 2016 QSM reconstruction challenge and data augmentation techniques (elastic transform, contrast change, and adding fake lesions). The local field were computed using the dipole kernel convolution to get the synthetic data pairs.

We apply domain-specific batch normalization layers in networks while allowing them to share all other model parameters. For synthetic data, L2 loss was used.

$$L_S = \left\| \chi_s - \chi_{ground-truth} \right\|_2$$

where $\chi_s$ is the network QSM output of synthetic data, and $\chi_{ground-truth}$ is the QSM label of synthetic data.

For real data, data consistency loss was used.

$$L_T = \left\| W_t(e^{jd*\chi_t} - e^{jf_t}) \right\|_2$$

where $W_t$ is the data weighting term to incorporate the noise weighting term and brain mask, $d$ is dipole kernel, $*$ is convolutional operator, $\chi_t$ is the network QSM outputs of real data, $f_t$ is local field of real data.

Total loss incorporates these two losses

$$L_{Total} = L_S + \lambda L_T$$

where $\lambda$ is loss weight which increases during training. The proposed method is denoted as QSMInvNet+.

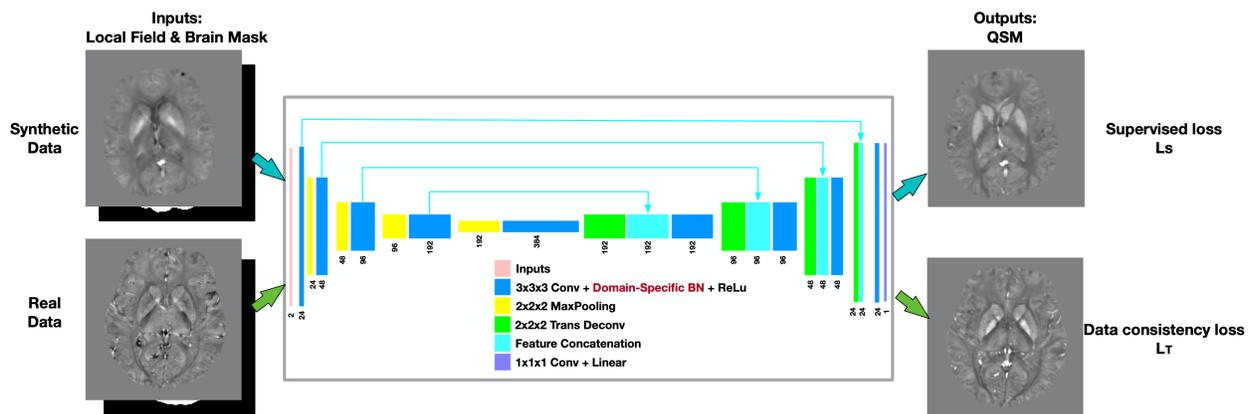

Fig 1. Neural network architecture of QSMInvNet+. It has an encoder-decoder structure with 9 convolutional layers (kernel size 3x3x3, same padding), 9x2 domain-specific batch normalization layers, 9 ReLU layers, 4 max pooling layers (pooling size 2x2x2, strides 2x2x2), 4 nearest-neighbor upsampling layers (size 2x2x2), 4 feature concatenations, and 1 convolutional layer (kernel size 1x1x1, linear activation). To address the unsupervised domain adaption on real data, domain-specific batch normalization layers are utilized while sharing all other model parameters.

[Data]
9 multi-orientation datasets were acquired with 5 head orientations and a 3D single-echo GRE scan with isotropic voxel size 1.0x1.0x1.0mm³ on 3T MRI scanners [1]. The data processing was the same as [9]. For network training, 500 synthetic data were generated then cropped to image patches with patch size 96x96x96. During training, the RESHARP local field, magnitude image and brain mask of 36 scans (9*4 tilted head positions) were randomly cropped with patch size 96x96x96. QSMInvNet+ was trained using patch-based network with patch size 96x96x96. Adam was used as optimizer with the learning rate of 10$^{-4}$. The network was implemented using TensorFlow and was trained on one NVIDIA K80 GPU. TKD [7], MEDI [8], COSMOS was performed to get QSM results of normal head position. In addition, DL-based techniques QSMInvNet (which first trained on synthetic data, then fine-tuned the trained model on real data according to the physical model), QSMInvNet+, and uQSM [9] were performed to compare the QSM results. Quantitative metrics such as NRMSE, HFEN, SSIM were calculated using COSMOS result as a reference.

100 clinical data were acquired using susceptibility-weighted angiography on a 3T MRI scanner (GE Healthcare MR750) with data acquisition parameters: in-plane data matrix 288x224, FOV 22 cm, slice thickness 3 mm, first TE 12.6 ms, echo spacing 4.1 ms, 7 echoes, TR 39.7 ms, pixel bandwidth 244 Hz, and total acquisition time of about 2 minutes. Complex multi-echo images were reconstructed from raw k-space data with reconstruction resolution 0.76x0.76x3.0mm$^3$. The brain masks were obtained using the SPM tool. RESHARP method with spherical mean radius 6mm were used for background field removal. For network training, simulated data were generated with voxel size 0.76x0.76x3.0mm$^3$ with the method as above. The network was trained with patch size 128x128x64. TKD, MEDI, QSMInvNet, and QSMInvNet+ were performed to compare the QSM results.

## Results

Fig. 2 displays the quantitative metrics of QSM results with COSMOS map as a reference. QSMInvNet+ achieved the best score in PSNR, NRMSE, and HFEN.

|  | pSNR (dB) | NRMSE (%) | HFEN (%) | SSIM (0-1) |
| --- | --- | --- | --- | --- |
| TKD | 43.4 ± 0.5 | 91.4 ± 6.7 | 72.9 ± 6.6 | 0.831 ± 0.016 |
| MEDI | 41.5 ± 0.6 | 113.8 ± 7.6 | 100.4 ± 9.1 | **0.902±0.016** |
| QSMInvNet | 44.8±0.5 | 78.0±6.1 | 65.8±5.5 | 0.878 ± 0.014 |
| QSMInvNet+ | **45.9±0.5** | **68.7±5.8** | **60.6±5.7** | 0.888 ± 0.017 |
| uQSM | 45.6±0.4 | 71.4±5.0 | 62.8±5.0 | 0.890 ± 0.015 |

Fig 2. Means and standard deviations of quantitative performance metrics of 5 reconstructed QSM images with COSMOS as a reference on 9 multi-orientation datasets. QSMInvNet+ achieved the best in PSNR, NRMSE, and HFEN.

Fig.3 compared QSM results of a multi-orientation dataset. Compared with conventional methods TKD and MEDI, DL-based methods display QSM with clearer details and less artifacts. Among DL-based methods, QSMInvNet+ results show fewer shading artifacts.

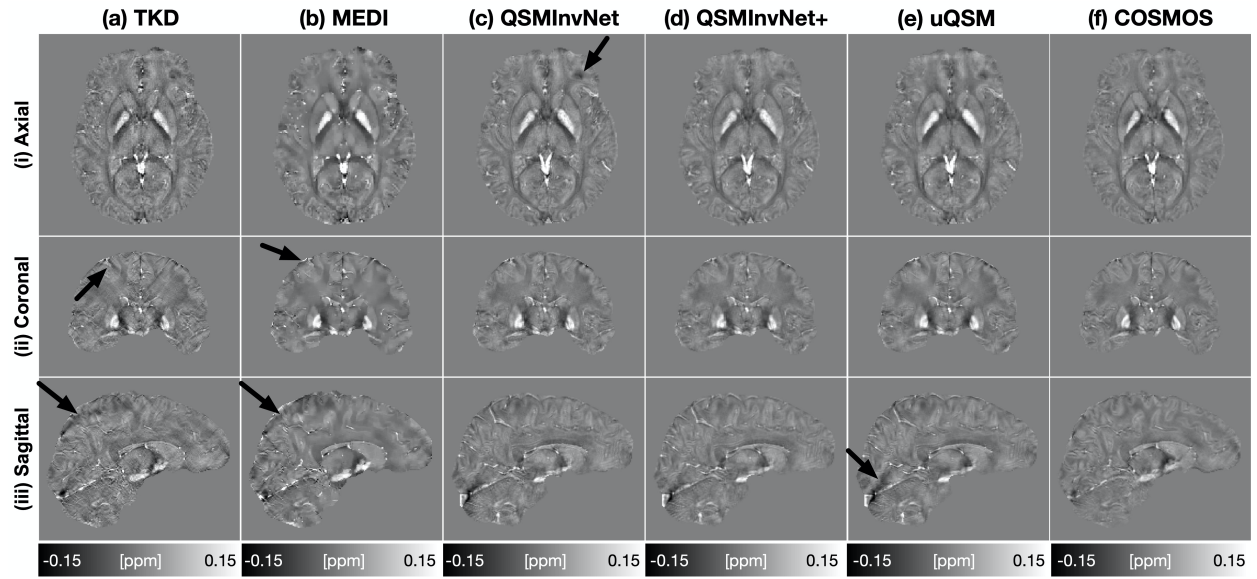

Fig 3. Comparison of QSM of a multi-orientation dataset. TDK (a), MEDI (b) and uQSM (e) results show black shading artifacts in the axial plane and streaking artifacts in the sagittal plane. QSMInvNet (c) displays better quality then TKD and MEDI but shows subtle artifacts. QSMInvNet+ (d) results have high-quality with clear details and invisible artifacts.

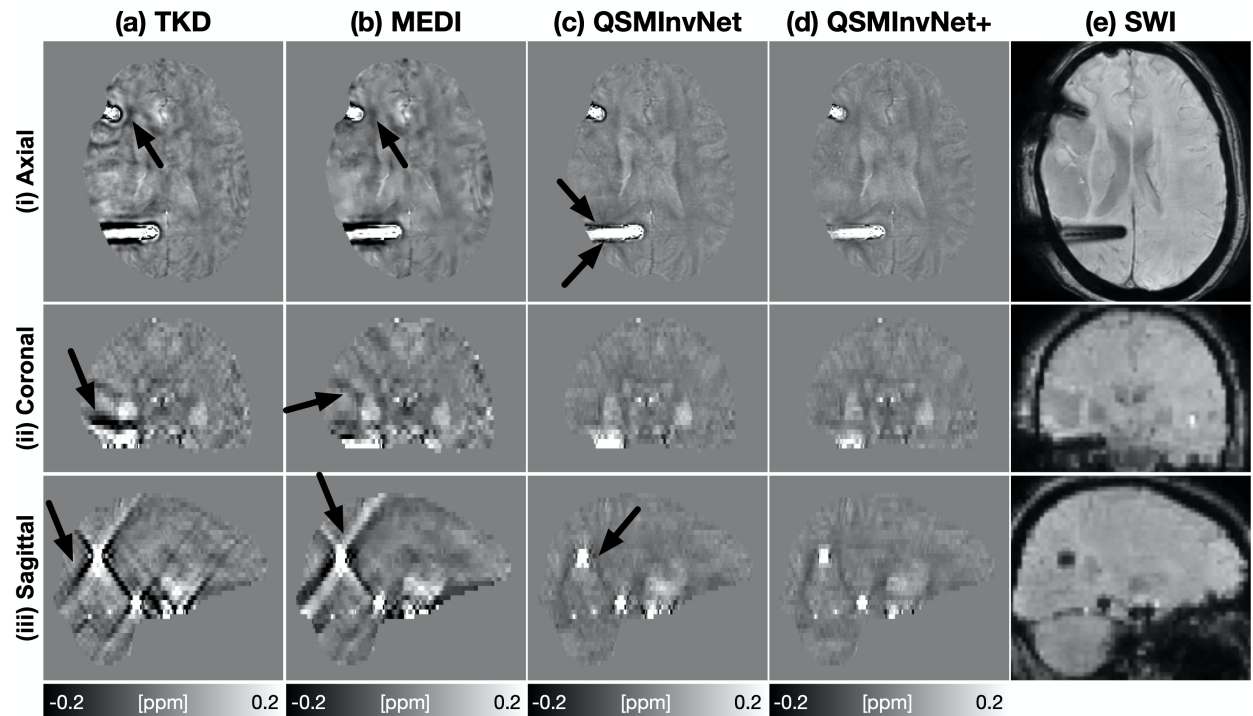

Fig 4. SM images (a-d) and SWI images (e) from a 37-year-old patient with surgical planning. TKD (a), and MEDI (b) maps showed over-smoothing and/or streaking artifacts. Compared with TKD and MEDI, QSMInvNet (c) and QSMInvNet+ (d) maps better preserve image details and show

subtle artifacts. QSMInvNet+ outperformed QSMInvNet with less shading artifacts around the hemorrhage regions (c, black arrows).

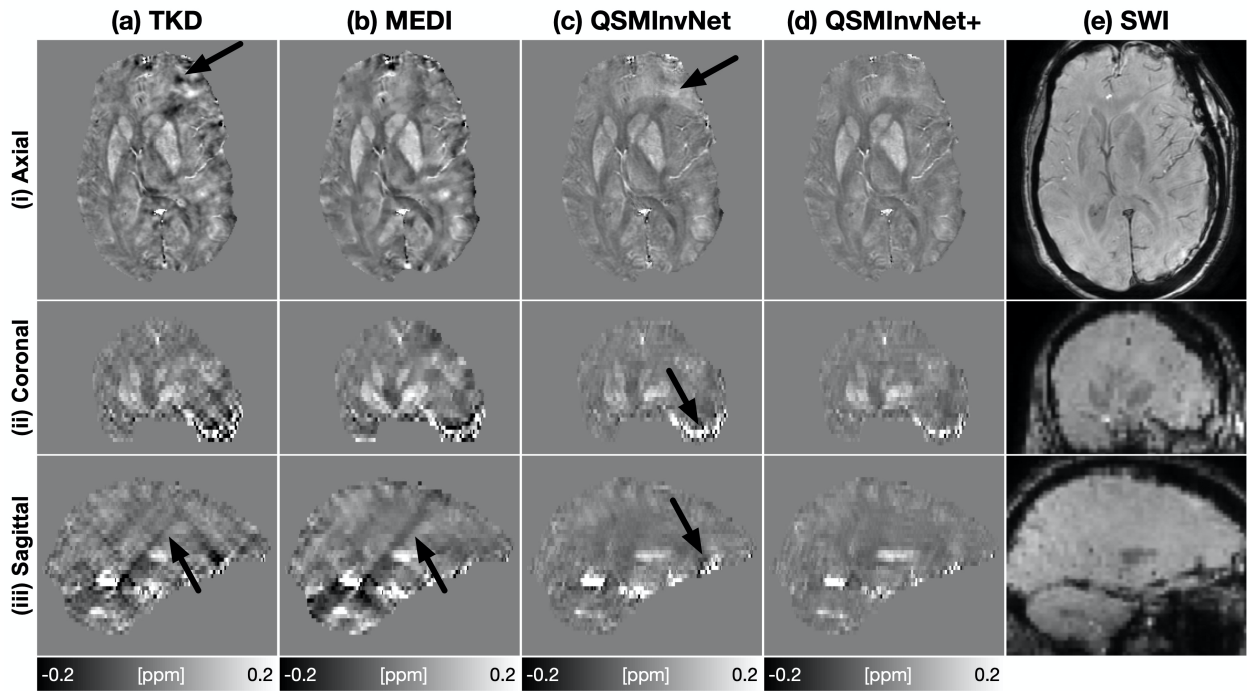

Fig 5. QSM images (a-d) and SWI images (e) from a 56-year-old subject with hemorrhagic intracranial metastases. TKD (a), and MEDI (b) maps showed over-smoothing and/or streaking artifacts. Compared with TKD and MEDI, QSMInvNet (c) and QSMInvNet+ (d) maps show less image artifacts and clearer details. QSMInvNet+ outperformed QSMInvNet with less artifacts especially around the hemorrhage regions (c, black arrows).

Fig. 4 and Fig. 5 compared QSM results of two clinical datasets with brain hemorrhage. TKD and MEDI results display sever streaking artifacts, while QSMInvNet and QSMInvNet+ outperformed with clear details and invisible artifacts. Compared with QSMInvNet, QSMInvNet+ images showed improved performance with less shading artifacts around the lesions.

## Discussion and Conclusion

We have proposed to apply domain-specific batch normalization to improve domain adaptation of neural networks learning from synthetic data. This method outperforms QSMInvNet - first learning from synthetic data then fine-tuning on real data. Since the method requires simulated data and minor changes on network architecture, it can be easily utilized to facilitate QSM research.

## Acknowledgements

We thank Professor Jongho Lee for sharing the multi-orientation QSM datasets.